\documentclass[prd,twocolumn,twoside,preprintnumbers,superscriptaddress,nofootinbib,%longbibliography
]{revtex4-1}

\usepackage{soul}
\usepackage{cancel}

\usepackage{amsmath,slashed}
\usepackage{graphicx,graphics,color}
\usepackage{dcolumn}
\usepackage{hyperref}
\usepackage{xspace}
\usepackage{enumerate}
\usepackage{varwidth}
\usepackage{physics}
\usepackage{amssymb}
\usepackage{mathtools}
\usepackage{dsfont}
\usepackage{booktabs,tabulary,longtable, multirow}
\hypersetup{
    colorlinks=true,
    linkcolor=blue,
    urlcolor=blue,
    citecolor=blue
}
\newcommand{\mpi}{M_\pi}
\newcommand{\mpin}{M_{\pi^0}}
\newcommand{\mpic}{M_{\pi^+}}

\newcommand{\beq}{\begin{equation}}
\newcommand{\eeq}{\end{equation}}
\newcommand{\diff}{\text{d}}

\newcommand{\Order}{\mathcal{O}}

\newcommand{\Mrho}{M_\rho}
\newcommand{\Grho}{\Gamma_\rho}

\newcommand{\GeV}{\,\text{GeV}}
\newcommand{\MeV}{\,\text{MeV}}

\renewcommand{\Im}{\text{Im}}

\newcommand{\disc}{\text{Disc}}

\newcommand{\BW}{\text{BW}}
\newcommand{\thr}{\text{thr}}

\allowdisplaybreaks[1]

\begin{document}

\title{Dispersive analysis of the $\boldsymbol{\phi \to \gamma \pi^0 \pi^0}$ process}
\author{Bai-Long Hoid}
\affiliation{Institut f\"ur Kernphysik and PRISMA$^{++}$  Cluster of Excellence, Johannes Gutenberg-Universit\"at Mainz,  55099 Mainz, Germany}  
\author{Igor Danilkin}
\affiliation{Institut f\"ur Kernphysik and PRISMA$^{++}$  Cluster of Excellence, Johannes Gutenberg-Universit\"at Mainz,  55099 Mainz, Germany}  
\author{Marc Vanderhaeghen}
\affiliation{Institut f\"ur Kernphysik and PRISMA$^{++}$  Cluster of Excellence, Johannes Gutenberg-Universit\"at Mainz,  55099 Mainz, Germany}

\begin{abstract} 
We present an analysis of the radiative decay $\phi \to \gamma \pi^0 \pi^0$ in a dispersive framework, where  the two-pion subsystem undergoes strong final-state interactions that cover the $f_0(500)$ and $f_0(980)$ regions. We employ a coupled-channel Muskhelishvili-Omn\`es framework that allows for a consistent treatment of two scalar resonances and crossed-channel singularities induced by the Born and vector-meson exchanges. We explicitly verify the equivalence between the modified and standard Muskhelishvili-Omn\`es representations for vector-meson pole contributions when the isoscalar Omn\`es matrix is chosen asymptotically bounded, and we adopt the standard representation in decay kinematics. This yields, for the first time, a parameter-free dispersive prediction for the kaon Born rescattering, which provides a dominant contribution. To obtain a good fit to the KLOE and SND data, we employ a once-subtracted coupled-channel dispersion relation with heavier left-hand cut contributions and two unknown subtraction constants. The results demonstrate the consistency among the data for $\pi\pi$ scattering,  $\gamma\gamma$ fusion, and $\phi $ radiative decay, thereby validating the underlying dispersive formalism and the input used for the hadronic Omn\`es matrix and left-hand cuts.  
\end{abstract}

\maketitle
%----------------------------------------------------------------------------------------
\section{Introduction}
%----------------------------------------------------------------------------------------
Radiative decays of light vector mesons into a photon and  two pseudoscalar mesons, $V\to\gamma P_1P_2$, provide a distinct probe of the light scalar sector of QCD~\cite{Klempt:2007cp}. Among them, measurements of the $\phi \to \gamma \pi^0 \pi^0$ and $\phi \to \gamma \pi^0 \eta$ channels were historically motivated by the expectation that sufficiently precise data could reveal the nature of the light scalar states $f_0(980)$ and $a_0(980)$~\cite{Achasov:1987ts}. It later turned out to be a challenging task due to the subtleties in extracting the relevant resonance couplings~\cite{Boglione:2003xh} and the observation that experimental information cannot clearly distinguish between molecular and compact structures of the scalars~\cite{Kalashnikova:2004ta}. Notwithstanding this controversy, radiative decays still provide additional insight into loosely constrained resonance properties of light scalar mesons beyond the hadron-hadron and two-photon scattering processes. In particular, for the $a_0(980)$ resonance, a dispersive analysis was performed for the related process $\phi \to \gamma \pi^0 \eta$, obtaining a reasonable description of the data with only two fit parameters~\cite{Moussallam:2021dpk}. 

In recent years, scattering amplitudes for photon-photon fusion into one or more mesons have been extensively analyzed in the context of the muon $g-2$~\cite{Muong-2:2006rrc,Muong-2:2021ojo,Muong-2:2023cdq,Muong-2:2025xyk,Aoyama:2020ynm,Aliberti:2025beg}, to provide an improved data-driven dispersive estimate of the contributions from hadronic light-by-light (HLbL) scattering~\cite{Colangelo:2015ama,Masjuan:2017tvw,Colangelo:2017fiz,Hoferichter:2018kwz,Eichmann:2019tjk,Bijnens:2019ghy,Leutgeb:2019gbz,Cappiello:2019hwh,Masjuan:2020jsf,Bijnens:2020xnl,Bijnens:2021jqo,Danilkin:2021icn,Stamen:2022uqh,Leutgeb:2022lqw,Hoferichter:2023tgp,Hoferichter:2024fsj,Estrada:2024cfy,Ludtke:2024ase,Deineka:2024mzt,Eichmann:2024glq,Bijnens:2024jgh,Hoferichter:2024bae,Holz:2024diw,Cappiello:2025fyf,Colangelo:2014qya}.  The leading singular topology of the HLbL tensor is generated by pseudoscalar poles, $P=\pi^0,\eta,\eta'$, which are described by their $P\to \gamma^* \gamma^*$ transition form factors~\cite{Hoferichter:2018dmo,Hoferichter:2018kwz,Holz:2024lom,Holz:2024diw}. After the pseudoscalar-pole contributions, the next relevant singularities arise from two-meson intermediate states, described by the subprocesses $\gamma^* \gamma^* \to P_1P_2$ within the dispersive framework in four-point kinematics. The pion-box contribution was first evaluated in Refs.~\cite{Colangelo:2017fiz,Colangelo:2017qdm}, together with the $\pi\pi$ $S$-wave rescattering effect that accounts model-independently for the contribution from the lowest scalar state $f_0(500)$. The rescattering formalism was later extended to cover the region of the $f_0(980)$ resonance~\cite{Danilkin:2021icn}, using a modified coupled-channel Muskhelishvili-Omn\`es representation~\cite{Danilkin:2019opj} with isoscalar hadronic-matrix input for $\pi\pi/K\bar{K}$ taken from  Ref.~\cite{Danilkin:2020pak}.  Due to the fundamental principles of analyticity, unitarity, and crossing, a reliable amplitude for $ \gamma^* \gamma\to\pi\pi$, used as input to HLbL scattering, is directly related to the radiative decay $\phi \to \gamma \pi^0 \pi^0$, in which the same $S$-wave rescattering effect dominates at low energies. This makes the decay a valuable testing ground for the underlying dispersive formalism and hadronic input. 
 
The radiative decay into neutral pion pairs $\phi \to \gamma \pi^0 \pi^0$ has been measured at several experimental facilities~\cite{SND:1998nxk,CMD-2:1999znb,Achasov:2000ym,KLOE:2002deh},  and, in contrast to $\phi \to \gamma \pi^+\pi^-$, is free from the initial- and final-state radiation contamination. Among these measurements, the $ \pi^0 \pi^0$ mass spectrum is dominated by the data from KLOE~\cite{KLOE:2002deh}, supplemented by the full data set from SND~\cite{Achasov:2000ym}. Additionally, a Dalitz-plot analysis was performed by KLOE for the $e^+e^- \to \gamma \pi^0 \pi^0$ events collected at a center-of-mass energy $\sqrt{s}\simeq M_\phi$  with about thirty times larger statistics~\cite{KLOE:2006vmv}. 
The latter, however, includes non-$\phi$ contributions (notably $e^{+}e^{-} \to \omega\pi^{0}$) and their interference. Since an unfolded, normalized Dalitz distribution is not available, we do not include Ref.~\cite{KLOE:2006vmv} and restrict ourselves to the unfolded $\pi^{0}\pi^{0}$ mass spectra from 
Refs.~\cite{KLOE:2002deh,Achasov:2000ym}. Apart from the kaon-loop model~\cite{Achasov:2001cj,Achasov:2005hm} applied in Ref.~\cite{KLOE:2006vmv} in a vector-meson-dominance form, this process was analyzed in the chiral unitary approach~\cite{Marco:1999df,Palomar:2003rb}, linear sigma model~\cite{Lucio:1999bb,Escribano:2006mb}, a formalism incorporating derivative interactions~\cite{Giacosa:2008st}, and a general decomposition of the underlying cross section~\cite{Isidori:2006we}. 

Since the $0^{++}$ scalar resonances $f_0(500)$ and $f_0(980)$ appear in the line shape, a consistent treatment of the coupled-channel $\pi\pi/K\bar K$ system is a prerequisite for performing a first dispersive analysis of the decay $\phi \to \gamma \pi^0 \pi^0$.  
The appropriate framework is provided by the
Muskhelishvili-Omn\`es (MO) representation, originally developed in the context of singular integral equations~\cite{Muskhelishvili1953,Omnes:1958hv} and now widely used to implement final-state interactions in dispersive amplitude analyses. In this approach, two-body unitarity determines the discontinuity of the partial waves along the right-hand cut, which is resummed by an Omn\`es function, or by an Omn\`es matrix in the coupled-channel case. The remaining singularities, in particular those generated by crossed-channel exchanges, enter through left-hand-cut (LHC) inhomogeneities, together with possible subtraction constants. This structure is directly suited to $\phi\to\gamma\pi^0\pi^0$, since the decay amplitude can be decomposed into $\pi\pi$ partial waves, while the isoscalar $S$-wave is strongly affected by $\pi\pi/K\bar K$ final-state interactions in the $f_0(500)$ and $f_0(980)$ regions. The Born terms and vector-meson exchanges provide the LHC input, whereas the soft-photon theorem imposes the corresponding kinematic constraint on the amplitude.

In this regard, two MO representations are available. Both include LHC input from the Born contribution and vector-meson exchanges. The modified approach~\cite{Garcia-Martin:2010kyn} requires only the unambiguous discontinuity of the vector-exchange contributions as input, but leads to numerically cumbersome integration contours in decay kinematics. The standard representation~\cite{Moussallam:2013una}, by contrast, involves only a direct integration above threshold, but is subject to a polynomial ambiguity in the treatment of vector exchanges. To address these issues, for the $S$-wave we propose an alternative approach in which only the unambiguous vector-meson pole contribution is implemented in the standard MO representation. We prove its equivalence to the modified MO approach provided the Omn\`es  matrix is chosen asymptotically bounded, as in Refs.~\cite{Danilkin:2020pak,Deineka:2024mzt}.

The remainder of the paper is structured as follows. After introducing the kinematics and helicity amplitudes for $\phi \to  \gamma \pi^0\pi^0$ in line with Refs.~\cite{Danilkin:2018qfn,Danilkin:2019opj}, we present the dispersive formalism in Sec.~\ref{sec:form} by first incorporating the soft-photon theorem~\cite{Low:1958sn}. We then describe the right-hand cut $\pi\pi/K\bar K$ $S$-wave dynamics by means of a two-channel Omn\`es matrix, the LHCs, and different MO representations. We fit the $ \pi^0 \pi^0$ line shape across the entire region and analyze the compatibility of the spectrum data and the dispersive constraints in Sec.~\ref{sec:results}. Our conclusions and outlook are presented in Sec.~\ref{sec:conc}.

%----------------------------------------------------------------------------------------
\section{Formalism}
\label{sec:form}
%----------------------------------------------------------------------------------------

%----------------------------------------------------------------------------------------
\subsection{Kinematics and helicity amplitudes}
\begin{figure}[t]
	\includegraphics[width=0.6\linewidth]{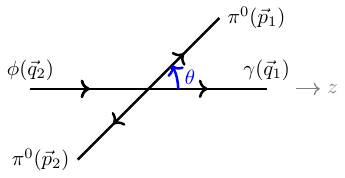}
	\caption{Momentum convention in the  $\pi^0 \pi^0$ center-of-mass frame.}
    \label{fig:kn}
\end{figure}

%----------------------------------------------------------------------------------------
We choose the momentum convention for the light meson decays as $V(q_2)\to\ \gamma(q_1)P_1(p_1) P_2(p_2)$. Accordingly, for the current process of interest, $\phi \to \gamma \pi^0 \pi^0$,  
\begin{equation}
    q_1^2=0,\quad  q_2^2\equiv q^2=M_\phi^2\, .
\end{equation}
The center-of-mass frame is chosen as in Fig.~\ref{fig:kn}, with the scattering angle $\theta=\angle (\vec{q}_1, \vec{p}_1)$. 
The Mandelstam variables are defined as
\begin{equation}
	s=(p_1+p_2)^2,\quad
	t=(p_1+q_1)^2,\quad
	u=(p_2+q_1)^2,
\end{equation}
and satisfy the relation $s+t+u= q^2 +2\mpin^2$. 

The hadronic tensor $H^{\mu\nu}$ is decomposed into a complete set of invariant amplitudes $F_i(s,t)$~\cite{Bardeen:1968ebo,Tarrach:1975tu,Drechsel:1997xv}, 
\begin{equation}
	H^{\mu\nu} = \sum_{i=1}^3F_i(s,t) L_i^{\mu\nu} \,,
\end{equation}
where explicit expressions for the tensors $L_i^{\mu\nu}$can be found, for example, in Ref.~\cite{Danilkin:2018qfn}. The corresponding helicity amplitudes are obtained by contracting the hadronic tensor with polarization vectors, 
\begin{equation}
    e_\gamma^{*\mu}(q_1,\lambda_1) e_{\phi}^\nu(q_2,\lambda_2) H_{\mu\nu} \equiv e^{i(\lambda_2-\lambda_1)\varphi}H_{\lambda_1 \lambda_2}(s,t)\,,
\end{equation}
where $\lambda_{1}$ and $\lambda_{2}$ denote the helicities of the photon and the $\phi$ meson, respectively. In total, three independent helicity amplitudes $H_{++}$, $H_{+-}$, and $H_{+0}$ contribute to the process. Assuming isospin conservation, only the isoscalar $(I=0)$ two-pion amplitude contributes to the decay $\phi \to \gamma \pi^0 \pi^0$. The isospin amplitude is therefore related to the charged- and neutral-pion amplitudes by
\begin{equation} \label{eq:isospin_pipi} 
H_{\lambda_1 \lambda_2}^{0}=-\sqrt{3}\,H_{\lambda_1 \lambda_2}^{c}=-\sqrt{3}\,H_{\lambda_1 \lambda_2}^{n}\, .
\end{equation}
Analogously for kaons, the helicity amplitudes with good isospin, denoted by $K^I_{\lambda_{1} \lambda_{2}}$, are related to the charged amplitudes as
\begin{equation}
     \binom{K_{\lambda_1 \lambda_2}^{0}}{K_{\lambda_1 \lambda_2}^{1}}=\left(\begin{array}{cc}-\sqrt{\frac{1}{2}} & -\sqrt{\frac{1}{2}} \\ -\sqrt{\frac{1}{2}} & \sqrt{\frac{1}{2}}\end{array}\right)\binom{K_{\lambda_1 \lambda_2}^{c}}{K_{\lambda_1 \lambda_2}^{n}}.
\end{equation}

%----------------------------------------------------------------------------------------
\subsection{Partial-wave dispersion relations}
%----------------------------------------------------------------------------------------
It is convenient to perform a partial-wave expansion of the helicity amplitudes~\cite{Jacob:1959at},
\begin{align}
	&H^I_{\lambda_{1} \lambda_{2}}\left(s, t \right)=\sum_{J}(2 J+1)\, h^I_{J, \lambda_{1} \lambda_{2}}(s)\, d_{\lambda_{2}-\lambda_{1}, 0}^{J}(\theta)\,, \nonumber\\
    &K^I_{\lambda_{1} \lambda_{2}}\left(s, t \right)=\frac{1}{\sqrt{2}}\sum_{J}(2 J+1)\, k^I_{J, \lambda_{1} \lambda_{2}}(s)\, d_{\lambda_{2}-\lambda_{1}, 0}^{J}(\theta)\,, 
\end{align}
where $d_{\lambda_{2}-\lambda_{1}, 0}^{J}(\theta)$ denotes the Wigner rotation function, and the normalization factor is chosen to ensure a consistent unitarity relation for identical and non-identical particles~\cite{Garcia-Martin:2010kyn}. 

Before setting up dispersion relations, one needs to identify the kinematic constraints of the helicity partial-wave amplitudes $h_{J, \lambda_{1} \lambda_{2}}(s)$ \cite{Danilkin:2018qfn}. For the dominant $S$-wave contribution, the only constraint comes from the soft-photon theorem~\cite{Low:1958sn}, which implies the behavior
\begin{align}\label{eq:soft-photon}
	 &\left.h^0_{0,++}(s)\right|_{s \rightarrow q^{2}}\simeq\Order\left(s-q^{2}\right),\notag\\
	&(k^0_{0,++}(s)-\left.k_{0,++}^{0, \text {Born}}(s))\right|_{s \rightarrow q^{2}}\simeq\Order\left(s-q^{2}\right)\,,
\end{align}
since the Born contribution  enters the $K\bar{K}$ channel only through the process $\phi\to \gamma K^+K^-$.  
After enforcing the kinematic constraints in Eq.~\eqref{eq:soft-photon}, one can formulate MO dispersion relations in two formally equivalent ways: either by  only subtracting  the Born term from the partial waves and treating the remaining LHC contributions through their discontinuities (the modified representation), or by incorporating the full LHC input explicitly in the inhomogeneity (the standard representation). In the following, we approximate the LHCs by the charged-kaon Born term and vector-meson exchanges; any remaining LHC contributions are assumed to be slowly varying in the physical region and are absorbed into the subtraction procedure. In an unsubtracted form for the coupled $\pi\pi/K\bar{K}_{I=0}$ system, the modified MO representation reads
\begin{align}
	\label{eq:unrepmod}
	\begin{pmatrix} h_{0,++}^{0}(s)\\
		k_{0,++}^{0}(s) \\ \end{pmatrix}&
	=\begin{pmatrix} 0\\
		k_{0,++}^{0,\text{Born}}(s) \\ \end{pmatrix}
	+\left(s-q^{2}\right) \bold{\Omega}_{0}^{0}(s) \notag\\
	\times \Bigg[-&\int_{4\mpi^2}^{\infty} \frac{\diff s'}{\pi} \frac{\Im\left(\bold{\Omega}_{0}^{0}\left(s'\right)^{-1}\right)}{(s'-s)(s'-q^{2})}\begin{pmatrix}0\\
		k_{0,++}^{0,\text{Born}}(s') \\ \end{pmatrix} \notag\\
	 +&\int_{\mathcal{C}_L} \frac{\diff z}{\pi} \frac{\bold{\Omega}_{0}^{0}\left(z\right)^{-1}}{(z-s)(z-q^{2})}\disc \begin{pmatrix} h_{0,++}^{0,V}(z)\\
			k_{0,++}^{0,V}(z) \\ \end{pmatrix}\Bigg],
\end{align}
where we set $\mpi=\mpic$ in the isospin limit, and ${\mathcal{C}_L} $ denotes the integration contour along the vector-exchange LHCs. The discontinuity of a general function $f(z)$ is defined as 
\begin{equation}
	\disc f(z)=\frac{1}{2i}\left(f(z+i\epsilon)-f(z-i\epsilon)\right).
\end{equation}
On the other hand, the standard MO representation is given by
\begin{align}
\label{eq:unrepsta}
	\begin{pmatrix} h_{0,++}^{0}(s)\\
		k_{0,++}^{0}(s) \\ \end{pmatrix}&
	=\begin{pmatrix} h_{0,++}^{0,V}(s)\\
		k_{0,++}^{0,L}(s) \\ \end{pmatrix}
	+\left(s-q^{2}\right) \bold{\Omega}_{0}^{0}(s) \notag\\
	\times\Bigg[-&\int_{4\mpi^2}^{\infty} \frac{\diff s'}{\pi} \frac{\Im\left(\bold{\Omega}_{0}^{0}\left(s'\right)^{-1}\right)}{(s'-s)(s'-q^{2})}\begin{pmatrix} h_{0,++}^{0,V}(s')\\
		k_{0,++}^{0,L}(s') \\ \end{pmatrix}\Bigg].
\end{align}
where, in contrast to the modified MO representation, the complete LHC contributions from Born and vector-meson exchanges are absorbed into the terms labeled by $L \equiv \text{Born}+V$. In both representations, the Omn\`es matrix encodes the hadronic scattering information and is given by
\begin{equation}
    \bold{\Omega}_{0}^{0}(s)=\left(\begin{array}{ll}\Omega_{0}^{0}(s)_{\pi\pi\to\pi\pi} & \Omega_{0}^{0}(s)_{\pi\pi\to K\bar{K}}  \\ \Omega_{0}^{0}(s)_{K\bar{K}\to\pi\pi}  & \Omega_{0}^{0}(s)_{K\bar{K}\to K\bar{K}} \end{array}\right).
\end{equation}
It is constructed from a data-driven $N/D$ analysis~\cite{Danilkin:2020pak}, where the fit is performed using the latest Roy and Roy-Steiner results for $\pi\pi\to \pi\pi$~\cite{Garcia-Martin:2011iqs} and $\pi\pi\to K\bar{K}$~\cite{Pelaez:2020gnd}, respectively. The crucial difference of the present work compared to other implementations is that it is asymptotically bounded, i.e., it satisfies a once-subtracted homogeneous equation,
\begin{equation}
\boldsymbol{\Omega}_0^0(s)
  =\boldsymbol{1}+\frac{s}{\pi} \int_{4M_\pi^2}^{\infty}
      \frac{\diff s'}{s'}\frac{\Im\,\boldsymbol{\Omega}_0^0(s')}{s'-s}\,,
\end{equation}
which is advantageous for low-energy applications, since it reduces the sensitivity of the MO dispersive integrals to high-energy input. Next, we specify the input from the LHCs in the following subsection.

%----------------------------------------------------------------------------------------
\subsection{Left-hand cuts}
%----------------------------------------------------------------------------------------
In order to proceed with the dispersive representation, one must specify the LHCs of the $\phi \to \gamma \pi^0 \pi^0$ amplitude (and analogously for the coupled $\phi \to \gamma K \bar K$ channels). In general, these LHCs originate from crossed-channel helicity amplitudes that exhibit singularities as functions of the Mandelstam variables $t$ and $u$. The leading singularity in the coupled channel is given by the charged-kaon pole (Born) term. Its contribution can be derived from scalar QED via the effective Lagrangian
\begin{align}
\mathcal{L}=&\,i\,g_{\phi K^+K^-} \phi^\mu 
\left(K^+(\partial_\mu K^-)-K^-(\partial_\mu K^+)\right)\nonumber\\
&+ 2\, e\, g_{\phi K^+K^-} \, \phi^\mu A_\mu K^+ K^-\,,
\end{align}
where  the coupling is related to the $\phi \rightarrow K^{+} K^{-}$ decay width by
\begin{equation}
	\Gamma_{\phi \rightarrow K^{+} K^{-}}=\frac{g_{\phi K^+K^-}^{2}}{48 \pi} M_{\phi}\left(1-\frac{4 M_{K^{+}}^{2}}{M_{\phi}^{2}}\right)^{3 / 2}.
\end{equation}
Numerically, we find $\abs{g_{\phi K^+K^-}}=4.51(2)$, based on the width from the Review of Particle Physics~\cite{ParticleDataGroup:2024cfk}. The partial-wave projection of the Born term then yields
\begin{align}
	\label{eq:Born}
	k^{c,\text{Born}}_{0,++}(s)&= e\,g_{\phi K^+K^-}\frac{4M_{K^+}^2 L_{K}(s)-2q^2}
	{s-q^2}\,,
\end{align}
with
\begin{align}
&L_{K}(s)=\frac{1}{\sigma_{K^+}(s)} \ln (\frac{1+\sigma_{K^+}(s)}{1-\sigma_{K^+}(s)})\,, \notag \\
&\sigma_{K^+}(s)=\sqrt{1-\frac{4M_{K^+}^2}{s}}\,.
\end{align}

The second class of LHCs is generated by vector-meson exchange diagrams. For the process $\phi\to\gamma\pi^0\pi^0$, the dominant contribution arises from $\rho$-exchange, via $\phi\to\rho\pi\to\gamma\pi^0\pi^0$. For completeness, we also consider the $K^*(892)$ contribution in the coupled channel $\phi\to K^*\bar K\to\gamma K\bar K$, although its numerical impact is found to be subdominant. We begin by analyzing these contributions within a Lagrangian-based approach and work initially in the narrow-width approximation. This allows us to capture the essential analytic structure of the vector-exchange amplitudes, while effects associated with finite widths are deferred to a later discussion. Using the interaction Lagrangian
\begin{equation}
	\mathcal{L}_{V P \gamma}=e\, C_{VP\gamma}\, \epsilon^{\mu \nu \alpha \beta}\, F_{\mu \nu}\, \partial_{\alpha}P\, V_{\beta}\,, 
\end{equation}
the coupling $C_{V P\gamma}$ is related to the decay width $V \rightarrow P\gamma$ via 
\begin{equation}
	\Gamma_{V \rightarrow P\gamma}=\frac{\alpha}{2}\, C_{V P\gamma}^{2} \frac{\left(M_{V}^{2}-M_{P}^{2}\right)^{3}}{3 M_{V}^{3}}\,.
\end{equation}
The relevant decay widths are taken from the PDG averages~\cite{ParticleDataGroup:2024cfk} and are summarized in Table~\ref{tab:couplings}, along with the corresponding radiative couplings. In the numerical evaluation for the $\rho$-exchange contribution, we use the branching ratio of $\rho^0\to \pi^0\gamma$ to obtain the isoscalar amplitude according to Eq.~\eqref{eq:isospin_pipi}.

\begin{table}[t]
	\renewcommand{\arraystretch}{1.3}
	\centering
	\begin{tabular}{l@{\hspace{1em}}l@{\hspace{1em}}l}
		\hline\hline
		Decay &Partial width  [$\text{keV}$] & Coupling [$\text{GeV}^{-1}$] \\\colrule
		$\rho^+\to\pi^+\gamma$ &$67.1(7.5) $ & $|C_{\rho^+\pi^+\gamma}| =0.362(20)$\\
		$\rho^0\to\pi^0\gamma$ & $69.3(11.8)$  & $|C_{\rho^0\pi^0\gamma}| =0.366(31)$\\
		$K^{*+}\to K^+\gamma$  & $50.4(4.7)$ & $|C_{K^{*+}K^+\gamma}| =0.418(19)$ \\
		$K^{*0}\to K^0\gamma$  & $116.4(10.0)$ & $|C_{K^{*0}K^0\gamma}| =0.635(27)$ \\\botrule
	\end{tabular}
	\renewcommand{\arraystretch}{1.0}
	\caption{The partial decay widths from PDG~\cite{ParticleDataGroup:2024cfk} and the extracted radiative couplings.}  
	\label{tab:couplings}
\end{table}

In a similar vein, we introduce the Lagrangian for the $VVP$-type interaction:
\begin{equation}
	\mathcal{L}_{\phi VP }=C_{\phi VP }\, \epsilon^{\mu \nu \alpha \beta}\,\partial_{\mu} \phi_\nu \,  \partial_{\alpha}P\, V_{\beta} \,. 
\end{equation}
For kinematically allowed channels, their couplings can be extracted from the corresponding partial widths, 
\begin{equation}
	\Gamma_{\phi \rightarrow VP}=\frac{C_{\phi VP }^{2}}{96 \pi M_{\phi}^{3}} \lambda^{3 / 2}\left(M_{\phi}^{2}, M_{V}^{2}, M_{P}^{2}\right),
\end{equation}	
with the K\"all\'en function
\begin{equation}
	\lambda(a,b,c)=a^2+b^2+c^2-2(a b + a c +b c)\,.
\end{equation}
The on-shell decays $\phi\to K^*\bar K$ are kinematically forbidden. Nevertheless, $K^*$ exchange still contributes off shell in the crossed channel and therefore generates a LHC. Numerically, we take 
\begin{equation}
    |C_{\phi K^{*} K}|\equiv |C_{\phi K^{*+} K^+}|=|C_{\phi K^{*0} K^0}|\approx 10 \GeV^{-1}\,,
\end{equation}
following  Ref.~\cite{Moussallam:2021dpk}, where these values were estimated by applying $U(3)$ flavor symmetry and large-$N_c$ arguments within resonance chiral theory~\cite{Ecker:1989yg,Prades:1993ys}. The coupling $C_{\phi\rho\pi}\equiv C_{\phi\rho^{\pm}\pi^{\pm}}=C_{\phi\rho^0\pi^0}$ is related to the decay process $\phi\to 3 \pi$. It gives $\abs{C_{\phi\rho\pi}}=1.18(2)\GeV^{-1}$ from the relation $\Gamma_{\phi\to3\pi}\approx 3\,\Gamma_{\phi \rightarrow \rho^0\pi^0}$ in the narrow-width limit, where the uncertainty is propagated from the corresponding branching ratio~\cite{ParticleDataGroup:2024cfk}. The impact of finite-width effects and of different parameterizations of the $\rho$ propagator on this value will be discussed in Sec.~\ref{sec:results}. The resulting vector-exchange contributions to the invariant amplitudes read
\begin{align}
	\label{eq:vamplitude}
	&F_1^{V}(s,t) =-\frac{e\,C_{VP\gamma} C_{\phi VP}}{4} \left(\frac{4\,t-q^2}{t-M_{V}^{2}}+(t\to u)\right), \notag \\
	&F_2^{V}(s,t) =\frac{e\,C_{VP\gamma} C_{\phi VP}}{4} \left(\frac{1}{t-M_{V}^{2}}+\frac{1}{u-M_{V}^{2}}\right), \notag \\
	&F_3^{V}(s,t) =\frac{e\,C_{VP\gamma} C_{\phi VP}}{2(t-u)}\left(\frac{1}{u-M_{V}^{2}}-\frac{1}{t-M_{V}^{2}}\right).
\end{align}
The $J=0$ partial wave can then be projected from the vector-exchange helicity amplitudes, 
\begin{align}
	\label{eq:v-PW}
	&h_{0,++}^{V}(s)=e\,C_{VP\gamma}C_{\phi VP}\Big\{\frac{L_V(s)}{\sigma_P(s)}
	\Big[-M_V^2 \notag\\
    &+q^2
	\Big(\frac{M_V^2-M_P^2}{s-q^2}\Big)^2\Big]
	+q^2\Big(\frac{1}{2}-\frac{M_V^2-M_P^2}{s-q^2}\Big)+(s-q^2)\Big\}\,,
\end{align}
where $V$ is either $\rho$ or $K^*$, which implies that $P$ is either $\pi$ or $K$ accordingly. The logarithmic function $L_V$ is given by
\begin{equation}
	\label{eq:log}
	L_V(s)=\ln(\frac{M_V^2-t_+(s)}{M_V^2-t_-(s)}),
\end{equation}
where
\begin{equation}
   t_\pm(s)=M_P^2+\frac{q^2-s}{2} \left(1\pm \sigma_P(s)\right). 
\end{equation}

We see that the vector-exchange invariant amplitudes~\eqref{eq:vamplitude} contain $t$- and $u$-dependent terms in the numerators. This off-shell dependence translates into a polynomial ambiguity in the $S$-wave amplitude: different Lagrangian realizations of vector-meson fields generate the last term in Eq.~\eqref{eq:v-PW} with opposite signs. The modified MO representation is therefore advantageous, as it is insensitive to this ambiguity and depends only on the discontinuity along the cut, albeit at the price of numerically cumbersome integration contours in decay kinematics. By contrast, the standard MO representation is affected by this ambiguity. We note, however, that the polynomial ambiguity can be eliminated in the standard MO representation by restricting to the pole part of vector exchange, i.e. by replacing $t$ and $u$ in the numerators of Eq.~\eqref{eq:vamplitude} with $M_V^2$.
This point is particularly important whenever one attempts to “isolate rescattering” contribution by subtracting a vector-meson-exchange contribution from the full amplitude. The resulting separation is representation independent only if the subtraction is performed at the level of the vector-meson pole part. Subtracting instead the full vector-meson-exchange contribution in the modified MO representation (as done, for example, in Ref.~\cite{Schafer:2023qtl}) requires a reshuffling of the subtraction polynomials within the MO representation.

In the following section, we demonstrate the exact equivalence of the $S$-wave results obtained with the two representations once the vector-pole prescription is adopted.

%----------------------------------------------------------------------------------------
\subsection{Analytic structure of the left-hand cuts}
%----------------------------------------------------------------------------------------
For space-like virtuality $q^2<0$, the singularities of the discussed LHCs appear only on the negative real axis. As a consequence, right- and left-hand cuts are well separated so that Watson's theorem~\cite{Watson:1954uc} is manifest in the elastic region. The situation changes when we move to the decay region with time-like virtuality---special attention needs to be paid to evaluating dispersive integrals in line with the underlying analytic structure. 

For the $k^{c,\text{Born}}_{0,++}$ amplitude, in addition to featuring a cut along the negative real axis inherited from the analytic structure of $L_{K}(s)$,
\begin{equation}
	\Im\, L_{K}(s)=\frac{\pi}{ \sigma_{K^+}(s)}\theta(-s)\,,
\end{equation}
it shows a pole singularity at $s=q^2$ as dictated by the soft-photon theorem (cf. Eq.~\eqref{eq:Born}).  Contrary to the scenarios of space-like photon virtualities~\cite{Colangelo:2015ama,Danilkin:2018qfn,Danilkin:2019opj,Hoferichter:2019nlq}, this pole sits on top of the unitarity cut in the kinematics of the $\phi$ radiative decay, in which $q^2>4\mpi^2$. Accordingly, a consistent implementation should be achieved via the analytic continuation $q^2\to q^2+i\epsilon$, a prescription that moves the pole away from the unitarity cut.

In contrast to the Born term, the vector-exchange amplitude $h_{0,++}^{V}$ vanishes for $s\to q^2$. Moreover, its singularity is determined by the function $L_V$,  which has two logarithmic branch points. Their positions can be determined from the condition that the argument of the function $L_V$ becomes negative, 
\begin{equation}
  \label{eq:con}
	 t_\pm(s)=M_V^2\,.
\end{equation}
Solving Eq.~\eqref{eq:con}  for the variable $s$, we find the branch points
\begin{align}
   s_\pm=& q^2 -\frac{M_V^2-M_P^2}{2M_V^2}
   \Big(q^2 +M_V^2\notag\\&-M_P^2  \pm \sqrt{\lambda(q^2,M_V^2, M_P^2)} \Big)\,.
\end{align}
To define the branch cuts, one notices that $M_V$ serves as an effective threshold for the $t$-channel singularities. Also, it is convenient to rewrite the K\"all\'en function as 
\begin{align}
    \lambda(q^2,M_V^2, M_P^2)&=\big(q^2-(M_V+M_P)^2\big)\notag\\&\times
    \big(q^2-(M_V-M_P)^2\big)\,.
\end{align}
Then, we determine three different situations (see also the discussion detailed in Ref.~\cite{Moussallam:2013una}): for $q^2\leq(M_V-M_P)^2$, both branch points $ s_\pm$ are real and negative. The structure of each cut can be determined by varying $M_V$, leading to two segments  $(-\infty,s_-]$ and $[s_+,0]$.  In the region $(M_V-M_P)^2< q^2<(M_V+M_P)^2$, the square root of  the K\"all\'en function generates an imaginary part and $ s_\pm$ become complex: the resulting cuts consist of the negative real axis as well as complex-plane contours. For $q^2\geq(M_V+M_P)^2$, $ s_\pm$ become real again, and both are located above the unitarity cut $4M_P^2$.

The decay kinematics of $\phi\to\rho\pi\to\gamma\pi^0\pi^0$ belongs to the third case, where the LHCs overlap with the unitarity cut. In detail, the $q^2+i\epsilon$ prescription renders the LHCs well separated from the unitarity cut, thereby allowing the application of the two MO representations without further contour deformations.

As both the Born and vector-exchange contributions generate cuts that overlap with the unitarity cut for the $\phi \to \gamma \pi^0 \pi^0$ process, Watson's theorem is violated so that the phase of the amplitude deviates from the $\pi\pi$ elastic scattering phase shift~\cite{Creutz:1969hv}.  Furthermore, applications of the MO representations should comply with the overlapping cut structure, which is realized differently in different approaches~\cite{Garcia-Martin:2010kyn,Moussallam:2013una}. In the modified representation, one should perform the dispersive integral along the LHCs that extend into the complex plane, with a consistently defined discontinuity that can be first tested for the $S$-wave amplitude~\eqref{eq:v-PW} itself before taking it as input to the complete representation~\eqref{eq:unrepmod}. Alternatively, for the standard approach, one should choose the correct form of the logarithmic function~\eqref{eq:log} above the unitarity threshold in Eq.~\eqref{eq:unrepsta}. This can be tested by comparison with a dispersive reconstruction of the triangle diagram~\cite{Moussallam:2013una,Hoferichter:2013ama,Baru:2020ywb}, where the discontinuity of the standard scalar loop function $C_0(s)$ turns out to be proportional to the logarithmic term in the $S$-wave projected vector-exchange amplitude~\eqref{eq:v-PW}. 

\begin{figure}[t]
	\includegraphics[width=\linewidth]{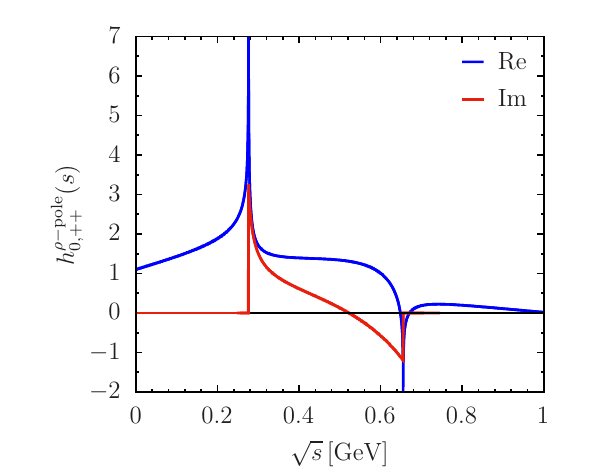}
	\caption{The real and imaginary parts of the $\rho$-pole LHC (couplings set to unity) from the explicit expression and from the dispersive representation~\eqref{eq:hv}. The integrals are evaluated along the complex cuts (see e.g., Ref.~\cite{Moussallam:2013una}). Two evaluations fall on top of each other, while the sharp behavior at $s_\pm$ is due to the narrow width.}
	\label{fig:V-test}
\end{figure}

With these specifics in mind, we continue to verify the equivalence between the two approaches for the vector-pole contributions.  For clarity, we work with the single-channel reduced case of the representations~\cite{Garcia-Martin:2010kyn,Moussallam:2013una}, while the generalization to coupled channels is straightforward.   It is easy to derive that both representations are equivalent if the following conditions are satisfied:
\begin{align}
	\frac{h_{0,++}^{V}(s)}{(s-q^2)\Omega_{0}^{0}(s)}
	=
	&\int_{4\mpi^2}^{\infty} \frac{\diff s'}{\pi} \frac{\Im\left(\Omega_{0}^{0}\left(s'\right)^{-1}\right)h_{0,++}^{V}(s')}{(s'-s)(s'-q^{2})} \notag\\
	+&\int_{\mathcal{C}_L} \frac{\diff z}{\pi} \frac{\Omega_{0}^{0}\left(z\right)^{-1}\disc\left(h_{0,++}^{V}(z)\right)}{(z-s)(z-q^{2})}\,.
\end{align}
For an Omn\`es matrix that behaves as $\Omega_{0}^{0}(s)\sim s^0$ for large $s$, this condition is exactly satisfied if we retain only the vector-pole contribution in the standard representation, since the pole piece itself fulfills
\begin{equation}
    \label{eq:hv}
	h_{0,++}^{V\text{-pole}}(s)=
	(s-q^2)\int_{\mathcal{C}_L} \frac{\diff z}{\pi} \frac{\disc\left(h_{0,++}^{V}(z)\right)}{(z-s)(z-q^{2})}\,.
\end{equation}
The closed-form expression for $h_{0,++}^{V\text{-pole}}(s)$ is obtained by dropping the term proportional to $(s-q^2)$ in  $h_{0,++}^{V}(s)$ given in Eq.~\eqref{eq:v-PW}. For Omn\`es matrices that asymptotically drop faster than constants, additional subtractions need to be introduced to ensure convergence of the dispersive integrals. One can verify that the equivalence still holds up to subtraction constants regardless of the number of subtractions. Therefore, this leads to a consistent definition of vector-pole contributions in the standard approach that is equivalent to the result in the modified method. Setting the couplings to unity, the $\rho$-pole contribution and its dispersive evaluation following Eq.~\eqref{eq:hv} for the kinematics of $\phi\to\gamma\pi^0\pi^0$ is shown in Fig.~\ref{fig:V-test}.

Normally, the finite widths of the exchange vector mesons have a limited effect on LHCs for space-like virtuality.  However, in the decay region, finite widths of exchange resonances  can no longer be ignored, especially when the resonance production becomes kinematically allowed for large enough $q^2$. This is exactly the case for the $\phi \to \gamma \pi^0 \pi^0$ process, where the decay $\phi\to\rho\pi$ is physically allowed as a two-body effective description of the $\phi\to3\pi$ decay process~\cite{Niecknig:2012sj, Danilkin:2014cra, Garcia-Lorenzo:2025uzc}. 

A simple solution to this issue is to introduce the widths of vector mesons through the modification $M_V\to M_V-i\Gamma_V/2$, which also separates the overlapping cuts.  More generally, to preserve a better analytic behavior of resonance propagators, one can resort to dispersively improved variants of  Breit-Wigner (BW) parameterizations~\cite{Zanke:2021wiq,Schafer:2023qtl,Messerli:2025rnv}.  We follow this prescription and apply an energy-dependent width to the $\rho$ propagator,
\begin{align}
	\Grho(t) 
	&= \theta(t - 4\mpi^2) \frac{\gamma_{\rho\to\pi\pi}(t)}{\gamma_{\rho\to\pi\pi}(\Mrho^2)} f(t)\, \Grho\,, \notag \\
	\gamma_{\rho \to \pi \pi}(t) 
	&= \frac{(t - 4\mpi^2)^{3/2}}{t}\,,
\end{align}
where $\Grho$ is the total width of the $\rho$ meson. The introduction of the barrier factor~\cite{VonHippel:1972fg,COMPASS:2015gxz},
\begin{equation}
    f(t)=\frac{\sqrt{t}}{\Mrho} \frac{\Mrho^2-4\mpi^2+4p_R^2}{t-4\mpi^2+4p_R^2}, \quad p_R = 202.4 \MeV\,,
\end{equation}
ensures better convergence of the integral for the dispersive $\rho$ propagator, 
\begin{align}
	P_\rho^\text{disp}(t) 
	&=  \int_{4\mpi^2}^\infty \frac{\diff x}{\pi} \frac{\Im\,P_\rho^\BW(x)}{x -t}\,, \notag \\
	\Im\,P_\rho^\BW(x) 
	&= N_\rho\frac{\sqrt{x} \, \Gamma_\rho(x)}{(M_\rho^2-x)^2 + x \Gamma_\rho(x)^2}\,,
    \label{eq:rho-disp}
\end{align}
where $4\mpi^2$ is the threshold for the $\rho \to \pi \pi$ decay process, and the factor $N_\rho$ is introduced to ensure the normalization 
\begin{equation}
	\int_{4\mpi^2}^\infty\frac{\diff x}{\pi}\,\Im\,P_\rho^\BW(x)=1\,.
\end{equation} 
In practice, an integration cutoff $\sqrt{s_\text{cut}}=2.0(2)\GeV$ is introduced to avoid unphysical higher-energy contributions, and its variation is taken as a source of systematic uncertainty.  Finally, we also consider a spectral form determined by the $P$-wave Omn\`es function $\Omega^1_1(s)$ to account for $\pi\pi$ final-state interactions, using the phase-shift input from Ref.~\cite{Caprini:2011ky}. By construction, it satisfies the dispersion relation
\begin{equation}
	\Omega^1_1(s)
	=  \int_{4\mpi^2}^\infty \frac{\diff x}{\pi} \frac{\Im\,\Omega^1_1(x)}{x -s}\,, 
\end{equation} 
with the integral to be evaluated with a cutoff, consistent with the dispersive BW form. 

To update the vector-exchange contributions in accordance with resonance widths, we observe that the propagator terms in the invariant amplitudes~\eqref{eq:vamplitude} should be replaced by the finite-width $\rho$ propagators, and Eq.~\eqref{eq:con} should be modified as follows:
\begin{equation}
	t_\pm(s)=t,\quad t \ge t_\thr,
\end{equation} 
where $t_\thr$ is the threshold of the unitarity cut in the $t$-channel, e.g., $t_\thr = 4\mpi^2$ for the $\rho$-exchange contribution. The total contribution can then be obtained from a spectral integral; for instance, in the dispersively improved BW parameterization,
\begin{equation}
    h_{0,++}^{V,\text{disp}}(s)= \int_{t_{\thr}}^{\infty} \frac{\diff x}{\pi}\,\Im\,P_\rho^\BW(x)\,\, h_{0,++}^{V\text{-pole}}(s,x)\,,
\end{equation}
which convolutes the narrow-width amplitude of the resonance propagator (with effective mass $x$)  with its spectral form of the imaginary part. Here $h_{0,++}^{V\text{-pole}}(s,x)$ is defined as the pole part of the narrow-width expression in Eq.~\eqref{eq:v-PW}  with the replacement $M_V^2\to x$ everywhere.

%----------------------------------------------------------------------------------------
\section{Numerical results}
\label{sec:results}
%----------------------------------------------------------------------------------------

In this section, we present the numerical results of our analysis, including the unsubtracted prediction for the Born rescattering term, once-subtracted fits to the $\phi \to \gamma \pi^0\pi^0$ experimental data, and the estimate of the branching ratio. 

The differential branching fraction for $\phi \to \gamma \pi^0\pi^0$  is given in terms of the three helicity amplitudes as
\begin{align}
	&\frac{\diff\Gamma_{\phi\to\gamma\pi^0\pi^0}}
	{\Gamma_\phi\,\diff\sqrt{s}} = 
	\frac{(q^2- s) \sqrt{s-4M_{\pi^0}^2}}{{384\pi^3} q^3\Gamma_\phi}\times \frac{1}{2}{\int_{-1}^1}\diff z\,  \notag\\
	& \quad \Big(\big|H_{++}(s,z)\big|^2 +\big|H_{+-}(s,z)\big|^2+\big|H_{+0}(s,z)\big|^2 \Big),
\end{align}
where $z=\cos\theta$, $\Gamma_\phi$ is the total decay width of the $\phi$ meson, and $s$ is the invariant mass squared of the $\pi^0\pi^0$ subsystem. The dominant $S$-wave contribution is entirely contained in the $H_{++}$ amplitude, while the effects of higher partial waves are approximated by the tree-level $\rho$-pole contribution.\footnote{We also estimated the kaon Born rescattering contribution in the $D$ wave using the newly evaluated coupled-channel $D$-wave Omn\`es input of Ref.~\cite{Danilkin:2025kyo} and found it to be negligible over the full $\pi\pi$ spectrum.}

\begin{figure}[t]
	\includegraphics[width=\linewidth]{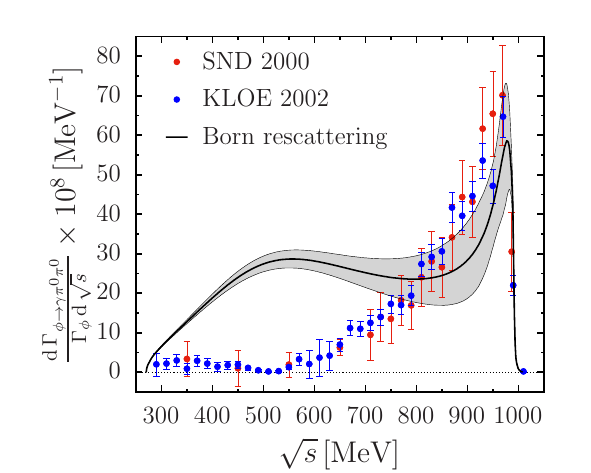}
    \includegraphics[width=\linewidth]{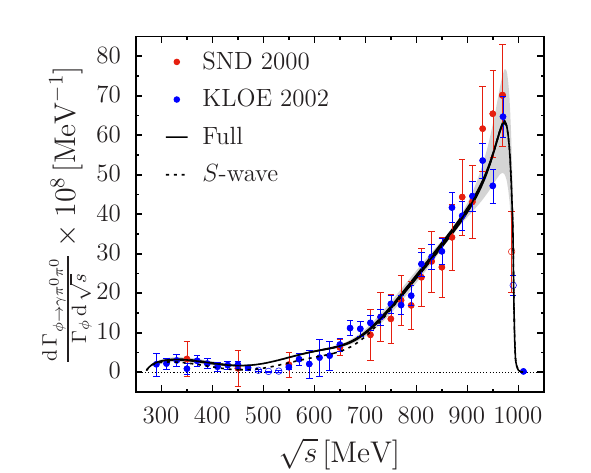}
	\caption{The differential branching ratio for $\phi\to\gamma\pi^0\pi^0$ as a function of the two-pion invariant mass $\sqrt{s}$. The top panel shows the prediction for the Born rescattering contribution. The bottom panel shows the final fit to the data: the black band represents the fit uncertainty alone, and the grey band denotes the total uncertainty. The $S$-wave contribution is indicated by the dashed line. Data points excluded from the fit are shown as open circles. }  
	\label{fig:fit}
\end{figure}

We first consider the unsubtracted representation and use it to obtain a parameter-free prediction for the Born-rescattering contribution. As in the case of $ \gamma^{(*)} \gamma^* \to\pi\pi$,  the kaon Born term leads to a rapidly convergent dispersive integral, whereas vector-pole contributions typically require additional subtractions for a stable numerical treatment. As shown in the top panel of Fig.~\ref{fig:fit}, the rescattering driven by the kaon-pole term provides a dominant contribution (with its uncertainty dominated by the Omn\`es input~\cite{Danilkin:2020pak}), largely reproducing the observed enhancement near the $f_0(980)$ region. However, the amplitude overestimates the strength in the low-energy region around $f_0(500)$, indicating the need for additional LHC contributions. 

To estimate the effect, we evaluate the sum-rule (SR) values of the Born contribution at $s=0$,
\begin{equation}
 	\label{eq:SR}
 	\begin{pmatrix}a^{\text{Born}}_{\text{SR}}\\
 	b^{\text{Born}}_{\text{SR}}\\ \end{pmatrix} =-\int_{4\mpi^2}^{\infty} \frac{\diff s' }{\pi}\frac{\Im\left(\bold{\Omega}_{0}^{0}\left(s'\right)^{-1}\right)}{s'(s'-q^{2})}\begin{pmatrix} 0\\
 		k_{0,++}^{0,\text{Born}}(s') \\ \end{pmatrix}.
\end{equation}
For the top and bottom rows, we obtain the central values 
\begin{eqnarray}
a^{\text{Born}}_{\text{SR}} &=&(-0.15 +0.12\,\mathrm{i})\GeV^{-2}, 
\nonumber \\
b^{\text{Born}}_{\text{SR}}&=&-0.77\GeV^{-2}. 
\end{eqnarray}
The sizable imaginary part of $a^{\text{Born}}_{\text{SR}}$ in the pion channel reflects a violation of Watson's theorem. If we instead determine these values from a fit to the data~\cite{KLOE:2002deh,Achasov:2000ym}, we find the shifts
\begin{eqnarray}
\Delta a^{\text{Born}}_{\text{fit}} &=&\big(0.34(1) -0.12(1)\,\mathrm{i}\big)\GeV^{-2},
\nonumber \\
\Delta b^{\text{Born}}_{\text{fit}}&=&-0.48(3)\GeV^{-2}. 
\end{eqnarray}
A strong violation of the sum rule with a sign change in the pion channel demonstrates the necessity of higher LHCs, in particular the $\rho$-pole contribution, which also generates imaginary parts below threshold.

To include vector-pole contributions as additional LHCs, we should fix the relative signs of the product of couplings entering the $\rho$ and $K^*$ poles to determine how these contributions interfere with the Born term (and with each other). In this regard, symmetry arguments are usually applied to fix the relative signs of the couplings in different contributions (see, e.g., Ref.~\cite{Moussallam:2021dpk} for $\phi \to \gamma \pi^0 \eta$). Instead, we relax the signs of the Born and vector-pole ($\rho$ and $K^*$) contributions and test the sensitivity of the data to the relative signs of these LHCs. The combination favored by the comparison between the unsubtracted representation and the data produces the same relative signs as those implied by the symmetry arguments.\footnote{Adopting $g_{\phi K^+K^-}>0$ as our convention, the products of couplings entering the left-hand cut contributions are $C_{\phi \rho \pi}\, C_{\rho \pi \gamma} < 0$, $C_{\phi K^{*0} K^0}\, C_{K^{*0} K^0 \gamma} < 0$,
and $C_{\phi K^{*+} K^+}\, C_{K^{*+} K^+ \gamma} > 0$.}

After fixing the relative signs, we introduce a once-subtracted representation to incorporate better convergent vector-pole contributions, 
 \begin{align}
 	\label{eq:onerep}
 	\begin{pmatrix} h_{0,++}^{0}(s)\\
 		k_{0,++}^{0}(s) \\ \end{pmatrix}
 	=&\begin{pmatrix} h_{0,++}^{0,\rho}(s)\\
 		k_{0,++}^{0,L}(s) \\ \end{pmatrix}
 	+\left(s-q^{2}\right) \bold{\Omega}_{0}^{0}(s)\Bigg[\begin{pmatrix}a\\
 	b \\ \end{pmatrix} \notag\\
 	-\frac{s-s_0}{\pi}\int_{4\mpi^2}^{\infty} &\frac{\diff s'}{s'-s_0} \frac{\Im\left(\bold{\Omega}_{0}^{0}\left(s'\right)^{-1}\right)}{(s'-s)(s'-q^{2})}\begin{pmatrix} h_{0,++}^{0,\rho}(s')\\
 		k_{0,++}^{0,L}(s') \\ \end{pmatrix}\Bigg],
 \end{align}
where $a$ and $b$ are two subtraction constants. In the following, we consider three implementations of the $\rho$-meson width in the pole term: a simple complex-mass prescription, a dispersively improved BW representation of Eq.~\eqref{eq:rho-disp}, and a $P$-wave Omn\`es function $\Omega^1_1(s)$. Consistently for each choice, the coupling $C_{\phi\rho\pi}$ is extracted from the $\phi\to3\pi$ decay width, yielding $0.84(1)$, $0.79(1)$, and $0.80(1)$ (all in $\GeV^{-1}$), respectively. These values supersede the narrow-width estimate quoted in Sec.~\ref{sec:form}. The impact of these implementations will be assessed in our fits.

To proceed,  we note that a bin-by-bin reweighting procedure should be considered because of its importance for radiative experiments. The KLOE analysis chose a universal bin width of $20\MeV$~\cite{KLOE:2002deh}, while SND used a wider bin width of $100\MeV$ for energies below $700\MeV$~\cite{Achasov:2000ym}. The finite bin-width effect can be taken into account by performing a bin average
\begin{equation}
	\frac{1}{\Delta{E_i}}\int_{E_i-\Delta{E_i}/2}^{E_i+\Delta{E_i}/2} \diff\sqrt{s}\times
	\frac{\diff\Gamma_{\phi\to\gamma\pi^0\pi^0}} {\Gamma_\phi\,\diff\sqrt{s}}
\end{equation} 
in the minimization. In practice, we implement the bin reweighting using a numerically efficient iterative solution~\cite{Ball:2009qv,Hoferichter:2019gzf,Hoferichter:2023bjm,Hoferichter:2025lcz}, which is shown to be equivalent to a bin-averaging correction. 

Taking into account finite bin widths,  we then test the compatibility of the KLOE~\cite{KLOE:2002deh} and  SND~\cite{Achasov:2000ym} data with our dispersive representation, choosing $s_0=0$ without loss of generality. We find that five data points cannot be described in a statistically acceptable way in our dispersive approach, even when allowing for two complex subtraction constants or a twice-subtracted representation.  Among them,  three energy bins in the range 480-540$\MeV$ dictate the observed inconclusive dip structure~\cite{Klempt:2007cp,KLOE:2002deh}. The mass spectrum at these energies yields an almost vanishing signal, while the associated uncertainties were estimated to be rather small. In this part of the spectrum, our dispersive description cannot reproduce the data within errors, yielding a contribution of $\chi^2 > 15$ from each bin.  In the measurement of Ref.~\cite{KLOE:2002deh},  the data below $650\MeV$ may be affected by an incomplete treatment of the coherent background from the $e^+e^-\to\omega\pi^0\to \gamma\pi^0\pi^0$ process~\cite{Achasov:2005hm}, which indicates underestimated uncertainties.\footnote{Given this situation of the low-energy data, we refrain from putting strong constraints on the amplitude at low energy. If we instead impose an Adler zero, the reduced chi-squared of the fit in Table~\ref{tab:fit} deteriorates to $\chi^2/\text{d.o.f.}\sim3.5$.} Furthermore, the two-kaon threshold region between $K^+K^-$ and  $K^0\bar{K^0}$ undergoes an enhancement of isospin-breaking effects~\cite{Achasov:1979xc,Hanhart:2007bd}. Since our hadronic input is constructed under the assumption of isospin symmetry, we refrain from including this region in our analysis. This removes one data point from KLOE~\cite{KLOE:2002deh} and one from SND~\cite{Achasov:2000ym}. 

 \begin{table}[t]
 	\renewcommand{\arraystretch}{1.3}
 	\centering
 	\begin{tabular}{l@{\hspace{2em}}r@{\hspace{2em}}r@{\hspace{2em}}r}
 	\hline\hline
 	Fits        & Fit 1         & Fit 2     & Fit 3         \\\hline
 	\multirow[t]{2}{*}{$\chi^2/\text{d.o.f.}$}
 	    & $68.0/49$     &  $55.9/49$    &  $61.6/49$         \\
 	  & $=1.39$            &   $=1.14$       &   $=1.26$      \\
 	$p$-value  &$3.7\%$        & $23.1\%$   & $10.7\%$    \\
 	$a\, [\GeV^{-2}]$ & $0.36(1)$    &  $0.36(1)$   &  $0.37(1)$  \\
 	$b\, [\GeV^{-2}]$ & $-1.72(4)$   & $-1.73(4)$  & $-1.75(4)$    \\ 
 	\botrule
 	\end{tabular}
      \renewcommand{\arraystretch}{1.0}
 	 \caption{Fits to the differential branching fraction for $\phi\to\gamma\pi^0\pi^0$. Fit 1 uses $M_\rho\to M_\rho-i\Gamma_\rho/2$; Fit 2 employs the dispersive propagator of Eq.~\eqref{eq:rho-disp}; Fit 3 uses the spectral representation derived from the $P$-wave Omn\`es function. Here d.o.f. denotes degrees of freedom.
     }  
 	\label{tab:fit}
 \end{table}

After excluding these problematic outliers, we find a good description of the two-pion mass spectrum using the once-subtracted representation~\eqref{eq:onerep}, with two real subtraction constants $a$ and $b$. The subtraction constants are fitted to the data and are shown in Table~\ref{tab:fit} for three different implementations of the finite width. As already observed in the original publication~\cite{KLOE:2002deh}, the fit is sensitive to the strength of the $\rho\pi$ term, and the spectrum prefers a smaller normalization. We find the same behavior, namely that a smaller $C_{\phi\rho\pi}$ coupling produces a better fit result, while the details of the spectral form only have a subdominant effect. We choose the best fit as our central result and show it in the bottom panel of Fig.~\ref{fig:fit}.  The differences among different fits are considered as a source of systematics, while the total systematic uncertainty is dominated by the coupled-channel Omn\`es input~\cite{Danilkin:2020pak}.  Finally, we find from our central fit
\begin{equation}    {\text{Br}({\phi\to\gamma\pi^0\pi^0}})=\frac{\Gamma_{\phi\to\gamma\pi^0\pi^0}} {\Gamma_\phi}=1.13(3)(5)\times10^{-4}\,,
\end{equation} 
where the first uncertainty is statistical (including the scale factor), and the second is systematic. The branching ratio is consistent with the PDG average $1.13(6)\times10^{-4}$~\cite{ParticleDataGroup:2024cfk}. 

%----------------------------------------------------------------------------------------
\section{Conclusions and outlook}
\label{sec:conc}
%----------------------------------------------------------------------------------------

We have presented a dispersive analysis of the radiative decay $\phi \to \gamma \pi^0 \pi^0$, using a coupled-channel Muskhelishvili-Omn\`es framework that satisfies analyticity and unitarity constraints.  By combining the $\pi\pi/K\bar{K}_{I=0}$ $S$-wave hadronic input with a consistent treatment of the left-hand cuts from the kaon pole and the $\rho$ and $K^*$ vector-meson poles, we have demonstrated the consistency between the data from hadronic scattering and the radiative decay $\phi \to \gamma\pi^0\pi^0$, and validated both the Omn\`es input and the dispersive formalism employed in this work and in $ \gamma^{(*)} \gamma^* \to\pi\pi$. The latter serves as key input for precision Monte-Carlo simulations of two-photon processes at $e^+e^-$ colliders, as implemented in HadroTOPS \cite{Lellmann:2025aje}.

As a first outcome of our analysis, we provided a parameter-free prediction for the kaon-pole contribution based on the unsubtracted representation. The charged-kaon left-hand cut derived from scalar QED and the effective Lagrangian completely fixes the $S$-wave Born amplitude $k^{c,\text{Born}}_{0,++}(s,q^2)$, once $g_{\phi K^+K^-}$ is determined from $\phi\to K^+K^-$.  This yields a dispersive estimate of the kaon-loop contribution commonly modeled in the literature.  The kaon pole already recovers the two-pion spectrum in the $f_0(980)$ region to a large extent, analogous to the traditional kaon-loop result, but with analyticity and coupled-channel unitarity imposed by construction.

As a further outcome of our analysis, we obtained a good description of the KLOE and SND spectra~\cite{Achasov:2000ym,KLOE:2002deh} with only two real  subtraction constants, suggesting that the combined effect of the kaon-pole left-hand cut and the vector-meson contributions already captures the dominant dynamics of the decay. In principle, the subtraction constants in our dispersive analysis could be further constrained by the higher-precision measurement of Ref.~\cite{KLOE:2006vmv}. However, its analysis was performed with a folded vector-meson-dominance description of the full Dalitz plot of $e^+e^-\to \gamma\pi^0\pi^0$, focusing on the interpretation of the nature of $f_0(980)$.  As a consequence, the published fit results are inherently model dependent and cannot be directly used for a dispersive treatment of the $\phi \to \gamma\pi^0\pi^0$ process. Nevertheless, provided that the Dalitz-plot data of Ref.~\cite{KLOE:2006vmv} are available, a combined dispersive treatment of $ \phi \to \gamma\pi^0\pi^0$ and the $\omega\to\pi^0\gamma^*$ transition form factor in the scattering region could offer a refined Dalitz analysis fulfilling analyticity and unitarity constraints.  

Another key ingredient of our analysis is the explicit verification of the equivalence between the modified and standard Muskhelishvili-Omn\`es representations for vector-meson pole contributions, once the pole prescription is chosen such that the dispersive integrals converge. This supports the use of the standard representation in decay kinematics, while removing the subtlety of an ambiguous polynomial term and retaining the underlying analytic structure with overlapping left- and right-hand cuts. Therefore, the alternative treatment in the current analysis can be straightforwardly extended to related radiative decays. In particular, the analysis of the channel $\phi \to \gamma\pi^0\eta$ provides an independent probe of the scalar-isovector sector~\cite{KLOE:2009ehb}. In that case, our demonstration of its equivalence to the modified approach in $\phi \to \gamma\pi^0\pi^0$ will be further tested in $\phi \to \gamma\pi^0\eta$, which has so far only been investigated within a modified representation~\cite{Moussallam:2021dpk}. Another natural extension concerns the heavy quarkonium decay $J/\psi \to \gamma\pi^0\pi^0$. The current study provides a test of the dispersive framework needed to treat such processes in decay kinematics, including the coupled-channel dynamics and crossed-channel singularities within available phase space. A detailed analysis of the low-energy spectra of $J/\psi \to \gamma\pi^0\pi^0$~\cite{BESIII:2015rug}  and related channels along these lines is in progress~\cite{Hoid:2026ehk}. 

 %----------------------------------------------------------------------------------------
 \begin{acknowledgments}
 Financial support by the DFG through the fund provided to the Research Unit  ``Photon-photon interactions in the Standard Model and beyond'' (Projektnummer 458854507 - FOR 5327) is gratefully acknowledged.
\end{acknowledgments}

\bibliography{ref}

\end{document}